\documentstyle[12pt]{article}

\def\slash#1{\rlap/#1}
\renewenvironment{thebibliography}[1]
 {\begin{list}{\arabic{enumi}.}
    {\usecounter{enumi} \setlength{\parsep}{0pt}
     \setlength{\itemsep}{0pt} \settowidth{\labelwidth}{#1.}
      \leftmargin\labelwidth\advance\leftmargin\labelsep\rightmargin=0pt
     \sloppy
    }}{\end{list}}

\begin{document}

\begin{titlepage}
\setcounter{page} {1}

\title{\centerline{\small\rm March 1993 \hfill DOE-ER\,40757-007}
\rightline{\small\rm CPP-93-07}   \bigskip
{\Large\bf NEUTRINO INTERACTIONS IN MATTER}\footnote{Talk given at the
{\em `Rencontres de Moriond'}, 30~January - 6~February 1993. To appear
in the Proceedings.}}

\author{\bf Palash B. Pal\\
\em Center for Particle Physics\\
\em Physics Department, University of Texas\\
\em         Austin, TX 78712, USA   \bigskip\bigskip \\
-- Electronic mail addresses -- \\
\sf \begin{tabular} {rl} BITNET: \qquad
&  phbd070\,@\,utxvms.bitnet \\
INTERNET: \qquad & phbd070\,@\,utxvms.cc.utexas.edu \\
DECNET: \qquad & 25643\,::\,phbd070
\end{tabular}}

\date{}
\maketitle
\vfill
\begin{quotation} \noindent
If a  fermion is travelling through a medium, it can have matter-induced
magnetic and electric dipole moments. These contributions conserve
chirality, and can be non-vanishing even for a Majorana neutrino.
Several implications for neutrino physics are discussed.
\end{quotation}
\vfill
\thispagestyle{empty}
\end{titlepage}
\normalsize

\section{Neutrino propagation in matter}
This subject beacame popular when
Wolfenstein calculated neutrino refractive index in matter, and
subsequently Mikheyev and Smirnov recognized resonant nature of
flavor oscillations triggered by matter effects. Application to the
solar neutrino problem is now standard \cite{palrev}.

To set up the stage, we outline a covariant calculation leading to
Wolfenstein's result. One
evaluates the self-energy function to determine the dispersion
relation, which can give the refractive index \cite{nr88,pp89,n89}.
Consider a massless neutrino for simplicity.
Chirality dictates that  its self-energy is of the form
	\begin{eqnarray}
\Sigma = R S L \,,
	\end{eqnarray}
where $R$ and $L$ are chirality projection operators.
In vacuum, the most general form of $S$ is
	\begin{eqnarray}
S = a (k^2) \slash k \,.
	\end{eqnarray}
Thus, the pole of the propagator is at
$k^2=0$, i.e., the particle is massless to all orders.

In a medium characterized by the center-of-mass velocity $v^\mu$,
	\begin{eqnarray}
S = a \slash k + b \slash v
	\end{eqnarray}
in general. Thus, the dispersion relation changes, which is
responsible for a refractive index different from unity.

\section{Electromagnetic vertex}
	\begin{figure}[h]
\begin{center}
\begin{picture}(50,24)(0,6)
\thicklines
\put(5,10){\line(1,0){40}}
\multiput(15,10)(20,0){2}{\vector(1,0){2}}
\put(25,10){\circle*{5}}
\multiput(25,13)(0,6){4}{\oval(3,3)[r]}
\multiput(25,16)(0,6){3}{\oval(3,3)[l]}
\put(10,15){\makebox(0,0){{\large$\nu (k)$}}}
\put(40,15){\makebox(0,0){{\large$\nu (k')$}}}
\put(34,25){\makebox(0,0){{\large$\gamma(q)$}}}
\end{picture}
\end{center} \vspace{-5mm}
\caption[]{\sf The effective vertex with photon.}\label{f:vertex}
	\end{figure}
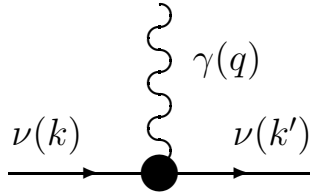
In general, the vertex in Fig.~\ref{f:vertex} can be written as
	\begin{eqnarray}
\bar u (k') \Gamma_\lambda (k,k',v) u (k) A^\lambda\,,
	\end{eqnarray}
where the vector $v^\mu$ has been defined earlier.
Conservation of charge implies the condition:
	\begin{eqnarray}
q^\lambda \Gamma_\lambda &=& 0 \,.
	\end{eqnarray}
In the vacuum where $v^\mu$ does not exist, the neutrino has no
charge, so that
	\begin{eqnarray}
\Gamma_\lambda (k,k,0) = 0 \,.
	\end{eqnarray}
These imply the following most general form
for $\Gamma_\lambda$ in the vacuum \cite{n82,kg83,k84}:
	\begin{eqnarray}
\Gamma_\lambda = (q^2 \gamma_\lambda - q_\lambda \slash q)
(R+r\gamma_5) + i \sigma_{\lambda\rho} q^\rho (D_M + D_E \gamma_5) \,.
	\end{eqnarray}

In matter, additional terms are possible because of the 4-vector $v$
\cite{np89,ss89}:
	\begin{eqnarray}
\Gamma_\lambda^\prime = i D_E^\prime (\gamma_\lambda v_\rho -
\gamma_\rho v_\lambda ) q^\rho \gamma_5
+ i D_M^\prime \epsilon_{\lambda\rho\alpha\beta} \gamma^\rho \gamma_5
q^\alpha v^\beta \,.
	\end{eqnarray}
It is easy to understand these terms in co-ordinate space if all form
factors are assumed to be momentum {\em in}dependent.
The vacuum part can then be written as
	\begin{eqnarray}
\bar \psi \gamma_\lambda \psi \partial_\rho F^{\lambda\rho}
(R+r\gamma_5) +
\bar \psi \sigma_{\lambda\rho} \psi F^{\lambda\rho}
(D_M + D_E \gamma_5) \,,
\label{vac-coord}
	\end{eqnarray}
whereas the extra terms in a medium are
	\begin{eqnarray}
D_E^\prime \bar \psi \gamma_\lambda \gamma_5 \psi \, v_\rho F^{\lambda\rho}
+ D_M^\prime \bar \psi \gamma_\lambda \gamma_5 \psi \,
v_\rho \tilde F^{\lambda\rho} \,.
\label{D'-coord}
	\end{eqnarray}
In the non-relativistic limit, since
$\bar \psi \gamma_0 \gamma_5 \psi \to  0$,
$\bar \psi \vec \gamma \gamma_5 \psi \to  \vec \sigma$,
in a frame where $v^\rho = (1, \vec0)$, we can write (\ref{D'-coord})
as
	\begin{eqnarray}
D_E^\prime \vec\sigma \cdot \vec E  +
D_M^\prime \vec\sigma \cdot \vec B \,.
	\end{eqnarray}
Hence, these are new contributions to dipole moments \cite{np89,ss89}.
Notice that these terms are chirality conserving, and
can be non-zero even for a Majorana neutrino \cite{np89}. Both these
properties are different from the vacuum dipole moment terms.

\section{Calculations at the leading order}
To the leading order in the Fermi constant \cite{dnp89,opss87},
	\begin{eqnarray}
\Gamma_\lambda &=& {\cal T}_{\lambda\rho} \gamma^\rho L \,, \\
{\cal T}_{\lambda\rho} &=& {\cal T}_T R_{\lambda\rho} +
{\cal T}_L Q_{\lambda\rho} + {\cal T}_P P_{\lambda\rho} \,,
	\end{eqnarray}
where, with $\tilde g_{\lambda\rho} = g_{\lambda\rho} - q_\lambda
q_\rho/q^2$ and $\tilde v_\lambda=\tilde g_{\lambda\rho} v^\rho$,
	\begin{eqnarray}
R_{\lambda\rho} &=&  \tilde g_{\lambda\rho} - Q_{\lambda\rho} \\
Q_{\lambda\rho} &=&  \tilde v_\lambda \tilde v_\rho / \tilde v ^2\\
P_{\lambda\rho} &=& i \epsilon_{\lambda\rho\alpha\beta} q^\alpha v^\beta /
\sqrt{(q \cdot v)^2 - q^2}\,.
	\end{eqnarray}
The form factors ${\cal T}_T$, ${\cal T}_L$, ${\cal T}_P$ have been
calculated in the leading order in $G_F$,
in the general case when the incoming and the outgoing
neutrinos in Fig.~\ref{f:vertex} may or may not be the same.
{}From this, one can obtain various physical effects, as
described below.

\paragraph*{Radiative decay~:}
In the vacuum, this is suppressed due to leptonic GIM.
A medium full of electrons and not muons or taons is not flavor
symmetric. Thus, the GIM mechanism is no more operative. The rates are
enhanced tremendously \cite{dnp90,gkl91}.

\paragraph*{Modification of forward scattering amplitude~:}
This occurs through the electromagnetic vertex, if the electron
scatters from the photon.
This was originally supposed to be large, as large as the Wolfenstein
term \cite{hor91}.
Later, it was pointed out that protons can also scatter from the photon,
and this cancels exactly the electron-scattering
contribution in the forward scattering amplitude for neutrinos
\cite{np92,sem92}.

\paragraph*{Induced electric charge~:}
Neutrinos in medium have a small charge induced by matter effects
\cite{os87},
$e^\nu_{\rm ind} = - (e G_F / \sqrt {2}) (1+ 4 \sin^2 \theta_W) (e^2
r_D^2)^{-1}$ where the Debye screening length is given by
$r_D^2=T/n_ee^2$ for a non-relativistic plasma
at temperature $T$. This gives $e^\nu_{\rm ind} \approx -2 \times
10^{-32} (1\,{\rm cm}/r_D)^2$. This is not measurable even for the
densest known plasma for which $r_D \simeq 10^{-4}$~cm.

\paragraph*{Plasmons~:}
Plasmons can decay into $\nu\bar\nu$. The rate has been calculated
\cite{dnp89}. Neutrinos can produce plasmons in a medium
\cite{opss87,saw92}.

\section{Outlook}
The effects may be large compared to the vacuum effects, but still not
hopeful for being observable. However, the subject is interesting, and
I am sure that the resonant neutrino oscillation is {\em not} the only
interesting physical effect. Maybe some other process,
like the Majoronic decay of neutrinos \cite{gkll92},
or other systems like the early universe \cite{BD91,EKM91}, will yield
some non-trivial effects. Or maybe in more non-trivial matter
distribution as in a crystal, some of the electromagnetic effects will
be enhanced enough to be observable.

I end by thanking my friend Jos\'e F Nieves, with whom I have done most
of the work on this subject. I also thank R Cowsik for some
discussions after my talk at the conference.

\end{document}